\documentclass[aps,prd,preprint,floatfix,nofootinbib,superscriptaddress]{revtex4}
\usepackage{graphicx,subfigure,epsfig,epsf}
\usepackage[utf8x]{inputenc}
\usepackage{amssymb,amsmath}
\usepackage{slashed}
\usepackage{feynmf}

\def\dnul{\partial_{\nu}}

\newcommand{\s}{\sigma}
\newcommand{\g}{\gamma}
\newcommand{\om}{\omega}

\newcommand{\beq}{\begin{equation}}
\newcommand{\eeq}{\end{equation}}
\newcommand{\bea}{\begin{eqnarray}}
\newcommand{\eea}{\end{eqnarray}}
\newcommand{\beas}{\begin{eqnarray*}}
\newcommand{\eeas}{\end{eqnarray*}}
\newcommand{\bcr}{\begin{center}}
\newcommand{\ecr}{\end{center}}
\newcommand{\vphi}{\langle \phi \rangle}

\def\Re{{\cal R \mskip-4mu \lower.1ex \hbox{\it e}\,}}
\def\Im{{\cal I \mskip-5mu \lower.1ex \hbox{\it m}\,}}
\def\ie{{\it i.e.}}
\def\eg{{\it e.g.}}

\def\etal{{\it et al.}}

\def\tev{\,{\ifmmode\mathrm {TeV}\else TeV\fi}}
\def\gev{\,{\ifmmode\mathrm {GeV}\else GeV\fi}}
\def\mev{\,{\ifmmode\mathrm {MeV}\else MeV\fi}}
\def\to{\rightarrow}

\begin{document}


\def\issue(#1,#2,#3){#1 (#3) #2} 
\def\APP(#1,#2,#3){Acta Phys.\ Polon.\ \issue(#1,#2,#3)}
\def\ARNPS(#1,#2,#3){Ann.\ Rev.\ Nucl.\ Part.\ Sci.\ \issue(#1,#2,#3)}
\def\CPC(#1,#2,#3){comp.\ Phys.\ comm.\ \issue(#1,#2,#3)}
\def\CIP(#1,#2,#3){comput.\ Phys.\ \issue(#1,#2,#3)}
\def\EPJC(#1,#2,#3){Eur.\ Phys.\ J.\ C\ \issue(#1,#2,#3)}
\def\EPJD(#1,#2,#3){Eur.\ Phys.\ J. Direct\ C\ \issue(#1,#2,#3)}
\def\IEEETNS(#1,#2,#3){IEEE Trans.\ Nucl.\ Sci.\ \issue(#1,#2,#3)}
\def\IJMP(#1,#2,#3){Int.\ J.\ Mod.\ Phys. \issue(#1,#2,#3)}
\def\JHEP(#1,#2,#3){J.\ High Energy Physics \issue(#1,#2,#3)}
\def\JPG(#1,#2,#3){J.\ Phys.\ G \issue(#1,#2,#3)}
\def\MPL(#1,#2,#3){Mod.\ Phys.\ Lett.\ \issue(#1,#2,#3)}
\def\NP(#1,#2,#3){Nucl.\ Phys.\ \issue(#1,#2,#3)}
\def\NIM(#1,#2,#3){Nucl.\ Instrum.\ Meth.\ \issue(#1,#2,#3)}
\def\PL(#1,#2,#3){Phys.\ Lett.\ \issue(#1,#2,#3)}
\def\PRD(#1,#2,#3){Phys.\ Rev.\ D \issue(#1,#2,#3)}
\def\PRL(#1,#2,#3){Phys.\ Rev.\ Lett.\ \issue(#1,#2,#3)}
\def\PTP(#1,#2,#3){Progs.\ Theo.\ Phys. \ \issue(#1,#2,#3)}
\def\RMP(#1,#2,#3){Rev.\ Mod.\ Phys.\ \issue(#1,#2,#3)}
\def\SJNP(#1,#2,#3){Sov.\ J. Nucl.\ Phys.\ \issue(#1,#2,#3)}



\bibliographystyle{revtex}


\title{Impact of a light stabilized radion in supernovae cooling} 



\author{Prasanta~Kumar~Das}
\email[]{Author(corresponding): pdas@goa.bits-pilani.ac.in}
\author{J.~R.~Selvaganapathy}
\author{Chandradew Sharma}
\author{Tarun Kumar Jha}
\author{V. Sunil Kumar}
\affiliation{Birla Institute of Technology and Science-Pilani, K. K. Birla Goa campus, NH-17B, Zuarinagar, Goa-403726, India }


\date{\today}

\begin{abstract} 
{In the Randall-Sundrum model where the Standard Model fields are confined to the TeV brane located at the orbifold point $\theta = \pi$ and the gravity peaks at the Planck brane located at $\theta = 0$, the stabilized modulus (radion) field is required to stabilize the size of the fifth spatial dimension. It can be produced copiously inside the supernova core due to nucleon-nucleon bremstrahlung, electron-positron and plasmon-plasmon annihilations, which then subsequently decays to  neutrino-antineutrino pair and take away the energy released in SN1987A explosion. Assuming that the supernovae cooling rate $\dot{\varepsilon} \le 7.288\times 10^{-27} \rm{GeV}$, we find the lower bound on the radion vev $\vphi  \sim  9.0$ TeV, $2.2$ TeV and $0.9$ TeV corresponding to the radion mass $m_\phi = 5$ GeV, $20$ GeV and $50$ GeV, respectively. } \\
{\bf Keywords:} Randall-Sundrum model, Radion, Supernovae cooling.
\end{abstract}

\maketitle


\section{Introduction }

Several new models based on extra spatial dimensions have been put forward to explain the large hierarchy between the Plack scale $M_{Pl}(\sim 10^{19}~ {\rm GeV}$) and the electro-weak scale $M_{EW}(\sim 100~ {\rm GeV})$ \cite{ADD}. Among them the Randall-Sundrum(RS) model is particularly interesting \cite{RS} since it solves  the hierarchy problem in an elegant manner.  According to this model the world is $5$-dimensional and the fifth spatial dimension is characterized by the angular coordinate $- \pi \le \theta \le \pi$. The space is an $S^1/Z_2$ orbifold (i.e. the point $(x,\theta)$ is identified with the point $(x,-\theta)$). The metric describing such a $5$-dimensional world is non-factorizable and a line element in this space-time is given by
\bea \label{eqn:rsmetric}
d s^2 = e^{-2 k R_c |\theta|}\eta_{\mu \nu} d x^\mu d x^\nu
- R_c^2 d \theta^2
\eea
\noindent
where $k$ is the bulk curvature constant and $x^\mu$ are the Lorentz coordinates of four dimensional surfaces of constant $\theta$. This theory postulates two $D_3$ branes along $x^\mu$ directions living in $5$-dimensional world: one is located at the orbifold point $\theta = 0$ where gravity peaks(strong) and the other at the orbifold point $\theta = \pi$ where the Standard Model(SM) fields reside and gravity is weak. The factor $e^{- 2 k R_c |\theta|}$ appearing in the metric is kown as the warp factor. 
 The compactification radius $R_c$($\sim$ the distance between the two $D_3$ branes) can be related to the vacuum expectation value (VEV) of the modulus field $T(x)$ which corresponds to the fluctuations of the metric over the background geometry given by $R_c$. Replacing $R_c$ by $T(x)$, we can rewrite 
the RS metric at the orbifold point $\theta = \pi$ as
\bea
d s^2 = g_{\mu \nu}^{vis} d x^\mu d x^\nu - T(x)^2 d \theta^2
\eea
\noindent 
where $g_{\mu \nu}^{vis} = e^{- 2 \pi k T(x)}\eta_{\mu \nu} 
= \left(\frac{\Phi(x)}{f}\right)^2 \eta_{\mu \nu}$. Here
$f^2 = \frac{24 M_5^3}{K}$ and  $M_5$ is the $5$-dimensional Planck scale. 
One is thus left with a scalar field $\phi(x)$ which is dubbed as the radion field 
\cite{GW}. The existence of the radion (modulus) field is a direct and straightforward consequence of the existence of the non-factorizable metric.  Randall and Sundrum \cite{RS} showed that if the above metric Eq.\ref{eqn:rsmetric} be a solution of the $5$-dimensional Einstein equations, then $k$ is to be related to the bulk cosmological constant and the vacuum energies of the two $D_3$ branes in a particular way \cite{RS}. The modulus field in the original RS model was massless and it had no potential: so it was not stabilized. One needs to generate a stable vacuum for $T(x)$ at $R_c$, which in turn can give
$\phi(x)$ a non-zero vev.  This is done  in the Goldberger-Wise
mechanism \cite{GW}, using a bulk scalar field with suitable interactions with the two $3$ branes, whereupon a potential for the modulus field is generated, and
one ends up with a radion of nonzero mass. In particular, the radion can be lighter than the other low-lying gravitonic degrees of freedom and can very well act as the first messenger of a scenario with compact extra dimensions, and reveal itself in collider experiments. Several studies on the observable implications of the radion are available in the literature \cite{Grasser}. 

 Beside the collider signals, the model predicts a variety of novel signals which can be tested in a class of astrophysical or cosmological observations. In particularly, the energy loss mechanism of the core-collapse SN1987A explosion and it's relevance in new physics context is an exciting area to work. A lot of studies in this direction have already been made and are available in the literature \cite{Raffelt}. It is interesting to see whether the light stabilized brane-world radion do have some role in the supernovae  cooling or not. The present work is intended to explore this possibility.    
\section{Supernova Explosion and Cooling}

Supernovae, the final state of an exploding star, come in two main observational varieties: Type II are those whose optical spectra exhibit Hydrogen lines and have less sharp peaks at maxima (of 1 billion solar luminosities), whereas the optical spectra for the Type I supernovae does not have any Hydrogen lines and it exhibits sharp maxima \cite{VHS}. Physically, there are two fundamental types of supernovae, based on what mechanism powers them: the thermonuclear supernovae and the core-collapse ones. Only supernovae Ia are thermonuclear type and the rest are formed by core-collapse of a massive star.  The core-collapse supernovae are the class of explosions which mark the evolutionary end of massive stars ($M \ge 8\,M_\odot$). The kinetic energy of the explosion carries about 1\% of the liberated gravitational binding energy of about $3\times10^{53}~{\rm ergs}$ and the remaining 99\% going into neutrinos. This powerful and detectable neutrino burst is the main astro-particle interest of the core-collapse supernovae.

 In the case of SN1987A, about $10^{53}~{\rm ergs}$ of gravitational binding energy was released in few seconds and the neutrino fluxes were measured by Kamiokande \cite{Kamio} and IMB \cite{IMB} collaborations. Numerical neutrino light curves can be compared with the SN1987A data where the measured energies are found to be ``too low''.  For example, the numerical simulation in \cite{Totani:1997vj} yields time-integrated values $\langle E_{\nu_e}\rangle\approx13~{\rm MeV}$, $\langle E_{\bar\nu_e}\rangle\approx16~{\rm MeV}$, and $\langle E_{\nu_x}\rangle\approx23~{\rm MeV}$.  On the other hand, the data imply 
$\langle E_{\bar\nu_e}\rangle=7.5~{\rm MeV}$ at Kamiokande and 11.1~MeV at IMB~\cite{Jegerlehner:1996kx}.  Even the 95\% confidence range for Kamiokande implies $\langle E_{\bar\nu_e}\rangle<12~{\rm MeV}$.  Flavor oscillations would increase the expected energies and thus enhance the discrepancy~\cite{Jegerlehner:1996kx}.  It has remained unclear if these and other anomalies of the SN1987A neutrino signal should be blamed on small-number statistics, or point to a serious problem with the SN models or the detectors, or is there a new physics happening in supernovae?

Since we have these measurements already at our disposal, now if we propose some novel channel through which the core of the supernova can lose energy, the luminosity in this channel should be low enough to preserve the agreement of neutrino observations with theory. That is ${\cal L}_{new\, channel} \le 10^{53}\, ergs\, s^{-1}.$
This idea was earlier used to put the strongest experimental upper bounds on the axion mass \cite{axions}. In large extra dimension scenario (where the weakness of $4-d$ gravity is obtained by the large size of the extra spatial dimensions via 
$M_{Pl}^2 = (2 \pi R)^d M_{D}^{d+2}$ \cite{ADD}),
the KK gravitons interact with the strength of ordinary gravitons and thus are not trapped in the supernovae core.  During the first few seconds after collapse, the core contains neutrons, protons, electrons, neutrinos and thermal photons(plasmons). There are a number of processes in which higher-dimensional gravitons can be produced. For the conditions that pertain in the core at this time (temperature $T \sim 30-70$ MeV, density $\rho \sim (3-10) \times 10^{14}$ g cm$^{-3}$), the relevant processes are shown below \\
(i) Graviton($\cal{G}$) production in Nucleon-Nucleon Brehmstrahlung: 
$N + N \to N + N + \cal{G}$ \\
(ii) Graviton production in photon fusion: 
$\gamma + \gamma \to \cal{G}$ \\
(iii) Graviton production in electron-positron annihilation process: 
$e^{-} + e^{+} \to \cal{G}$ \\
(iv) Graviton production in plasmon-plasmon(photon inside plasma becomes massive and called plasmon) annihilation process:
$\gamma_P + \gamma_P \to \cal{G}$ \\
 The constraint on luminosity of this process can be converted into a bound on the 4+d dimensional Planck scale $M_D$. 
Raffelt has proposed a simple analytic criterion based on detailed supernova simulations~\cite{Raffelt}: if any energy-loss mechanism has an emissivity greater than $10^{19}$ ergs g$^{-1}$ s$^{-1}$ then it will remove sufficient energy from the explosion to invalidate the current understanding of Type-II supernovae's neutrino signal. 

 The dominant process relevant for the SN1987A where the temperature is comparable to $m_\pi$ and so the strong interaction between N's is unsuppressed. This process can be represented as (see above)
\beq
 N + N \to N + N + \cal{G},
\eeq
where $N$ can be a neutron or a proton and $\cal{G}$ is a higher-dimensional graviton. 

\begin{center}
Table 1
\end{center}
\vspace*{-0.25in}
\begin{center}
\begin{tabular}{|c|c|c|}
\hline
Group/Collaboration & $M_D$ (GeV) & $d$   \\
\hline
Cullen \etal \cite{CP}  &  $\ge 50$ TeV,~ $\ge 4$ TeV,~$\ge 1$ TeV& 2,~3,~4 \\
\hline 
Barger \etal \cite{BHKZ}  &  $\ge 51$ TeV,~ $\ge 3.6$ TeV & 2,~3 \\
\hline 
Hannestad \etal \cite{HR}  &  $\ge 84$ TeV,~$\ge 7$ TeV & 2,~3 \\
\hline 
\end{tabular}
\end{center}
\noindent {\it Table 1: The lower bound on the higher dimensional Planck scale $M_D$ corresponding to the no. of extra spatial dimensions $d$ is shown.  The lower bound follows from the fact for any new physics channel contributing to the SN1987A energy loss, the loss rate $\dot{\varepsilon} \le 7.288\times 10^{-27} \rm{GeV}$.}

 For the core temperature $T=30$ MeV  and $\rho=3 \times 10^{14}$ g cm$^{-3}$, we list in Table 1 the results(lower bound on $M_D$) of various authors.  In addition it is worthwhile to mention that the KK gravitons produced in plasmon-plasmon collision which contributes in the supernovae cooling gives rise strong bound on $M_D$: for $d=2$ one finds $M_D \ge 22.5~{\rm TeV}$ and for $d=3$ one finds $M_D \ge 1.4~{\rm TeV}$ \cite{DSS}.

Another extra-dimensional model that can play a crucial role in this SN1987A cooling is the Randall-Sundrum(RS) model and was first looked at in \cite{MM}. The authors in \cite{MM} studied the impact of a light radion on neutrino-anti-neutrino oscillation. They found that for a light radion of mass $m_\phi \ge 1~{ {{\rm GeV}}}$ with $\vphi = 1~{\rm{TeV}}$, the interaction potential(arising due to the exchange of a radion between the supernoave matter and the neutrino-antoneutrino pair) does not affect the neutrino oscillation. However, the role of a light radion in the Supernovae cooling was not looked at. The present work is an effort in that direction. 

 We will see that how the radion produced in electron-positron, plasmon-plasmon annihilation takes part in the supernovae SN1987A cooling. 
\begin{figure}[htbp]
\centerline{\hspace{-3.3mm}
{\epsfxsize=10cm\epsfbox{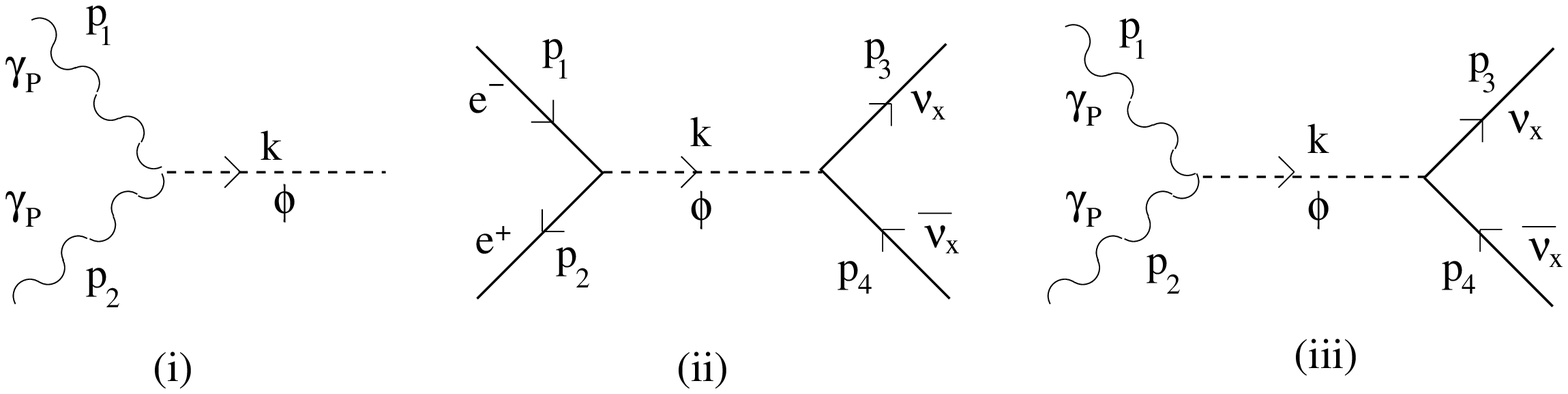}}}
\vspace*{-0.1in}
\caption{Feynman diagrams for the processes $\gamma_P + \gamma_P \to \phi$, $ e^- e^+ ~\stackrel{\phi}{\longrightarrow} ~ \nu_x \overline{\nu_x} $ and $\gamma_P + \gamma_P \to \phi \to \nu_x + \overline{\nu_x}$ (where $x = \mu,~\tau$) which contribute to the SN1987A cooling. }
\protect\label{feyn_radion_cooling}
\end{figure}

\noindent The three primary new physics diagrams that may contribute to the supernovae cooling are: 
(i) $\gamma_P + \gamma_P \to \phi$, ~(ii) $e^+ + e^- \to \phi \to \nu_x + \overline{\nu_x}$ and 
(iii)$\gamma_P + \gamma_P \to \phi \to \nu_x + \overline{\nu_x}$ where $x = \mu, \tau$. These are shown in Fig. \ref{feyn_radion_cooling}.
\section{Methodology of calculation}  
In Goldberger-Wise mechanism the modulus field $T(x)$ gains a mass and the radion field $\Phi$ ( where $\Phi = f e^{-\pi k T(x)} $ with $f = M_{Pl}$)couples to the brane matter through the trace of the energy-momentum tensor \cite{GW}
\beq
{\cal L}\ = \frac{\phi}{\vphi} T^{\mu}_{\mu}\
\label{interaction}
\eeq
where $\phi = \Phi - \vphi$ is the fluctuation of the radion field from the VEV 
$\vphi$. In the RS scenario, $\vphi \sim {\rm TeV}$ for $k R_c \simeq 12$ in order to produce the weak scale from the Planck scale through the exponental warp factor. 
Since radion coupling to SM matter is determined by the $4-d$ general covariance, it's coupling to matter is universal i.e. it couples to the trace $T_{\mu}^{\mu}$ of the energy-momentum tensor $T_{\mu\nu}$ of the SM matter fields which resides on the TeV brane and is given by 
\bea
T^\mu_\mu (SM) = \sum_{\psi} \left[\frac{3 i}{2} \left({\overline{\psi}}
\g_\mu \dnul \psi - \dnul{\overline{\psi}} \g_\mu \psi \right)\eta^{\mu\nu}
- 4 m_\psi {\overline{\psi}} \psi\right] - 2 m_W^2 W_\mu^+ W^{-\mu}
- m_Z^2 Z_\mu Z^\mu \nonumber \\
+ (2 m_h^2 h^2 - \partial_\mu h \partial^\mu h) + \cdots
\eea
Inside the supernovae, the relevant matter fields are the nucleons, electrons-positrons, plasmons and the interaction of braneworld radion ($\phi$) with them is given in Eq. \ref{interaction}. 

 Now for a generic $2\to N$ body scattering, the scattering cross section is given by
\bea
\sigma= \frac{1}{Flux} \int \prod_f \frac{d^3 p_f}{(2 \pi)^3 2 E_f} (2 \pi)^4 \delta^4 \left( p_1+p_2 -\sum_f p_f\right) \overline{|{\cal{M}}_{fi}|^2}
\eea 
where $Flux= 4 E_1 E_2 \upsilon_{rel}$. Here $E_1$, $E_2$ are the energies of the incoming  particles 1 and 2 whose masses are $m_1$ and $m_2$, respectively and $\upsilon_{rel}$ is the relative velocity between them. Defining the energy loss per unit mass  $ \dot{\varepsilon} = \frac{Q}{\rho_{SN}} \hspace{0.02in}(erg \hspace{0.02in} g^{-1} \hspace{0.02in} s^{-1} )$ for a generic $2 \to N$ scattering contributing to the cooling process at temperature $T$ one finds 
\begin{eqnarray}
\dot{\varepsilon} 
 & = & \frac{1}{\rho_{SN}}\prod^{2}_{i=1} \int \frac{ d^{3}\textbf{p}_{i} }{2E_{i}(2 \pi)^{3}} 
f_i(E_i)  \prod^{N}_{j=3} \int \frac{ d^{3}\textbf{p}_{j} }{2E_{j}(2 \pi)^{3}} (1 \pm f_j (E_j)) (2 \pi)^{4} \delta^{4}
\left(\sum_{i=1}^{2}p_i - \sum_{j=3}^{N} p_j\right)~\frac{1}{4} \sum_{spins} |M|^{2} \nonumber \\
\end{eqnarray}
where $f_i$, the occupation numbers for the initial colliding paricles (i.e. electrons, positrons, plasmons or nucleons) and $(1 \pm f_j)$, the Pauli blocking factors for the final state bosons ($+$ sign) and fermions($-$ sign)
 
\noindent For a general reaction of the kind $a + b \to c$, the above expression takes 
the form
\beq 
\sigma = \frac{1}{Flux} \overline{|{\cal{M}}_{fi}|^2} 2 \pi \delta(S - m_c^2).
\label{2to1crossgen}
\eeq

\noindent In the center of mass(c.o.m) frame, we use the notation $\sqrt{S}$ for the total initial energy
\bea
\sqrt{S}=E_1+E_2\label{cmenergy}\\
Flux = 4 E_1 E_2 \upsilon_{rel}=4 |\mathbf{p}| \sqrt{S},\label{relvel}
\eea 
where $|\mathbf{p}|=|\mathbf{p}_1|= |\mathbf{p}_2|= \frac{\lambda^{1/2}(S,m_1^2,m_2^2)}{2 \sqrt{S}}$ and $E_1$ and $E_2$ are the energies of the particles $1$ and $2$. The function $\lambda(x,y,z)(=x^2 + y^2 + z^2 - 2 x y -2 y z - 2 z x)$, is the standard  $K{\ddot{a}}$llen function.

~ Now we are concerned with the energy loss to radion. It is convenient and standard
\cite{Kolb,Raffelt} to define the quantities $\dot{\varepsilon}_{a+b \to c}$
which is the rate at which energy is lost to radion via the
process $a + b \to c $ where $c$ has a decay width. In terms of the cross-section 
$\sigma_{a+b \to c}$ the number densities  $n_{a,b}$ for a,b and the mass density
$\rho$, $\dot{\varepsilon}$ is given by
\beq
\dot{\varepsilon}_{a + b \to c.} = \frac{\langle n_a n_b
\sigma_{(a+b \to c)} v_{rel} E_{cm} \rangle}{\rho}
\label{emrate}
\eeq 
where the brackets indicate the thermal averaging and $E_{cm}(=E_a + E_b)$ is the center-of-mass(c.o.m) energy of the two colliding particles $a$ and $b$.
Note that in the present case, the final state radion has a smaller decay width but is  stable over the size of the neutron star because of it's large life time $\sim 10^9 (100 ~MeV/m)^3$ yr (See \cite{Han:1999sg}) and thus it can escape the supernovae while allowing it to cool.

 Next we are interested in the processes in which electron-positron and plasmon-plasmon collisions produce radion followed by it's decay to muon and tau neutrino anti-neutrino pair. These $2 \to 2$ (i.e. $a + b \to c \to d + g$) scatterings will also contribute in the overall SN1987A cooling. The volume energy-loss rate ($\dot{\varepsilon}$) per unit mass of the supernovae matter can be calculated as 
\begin{eqnarray}
\dot{\varepsilon} & = & \frac{1}{\rho_{SN}} \prod^{2}_{i=1} \int \frac{ d^{3}\textbf{p}_{i} }{2E_{i}(2 \pi)^{3}} 
f_i(E_i)  \prod^{4}_{j=3} \int \frac{ d^{3}\textbf{p}_{j} }{2E_{j}(2 \pi)^{3}} (1 \pm f_j (E_j)) (2 \pi)^{4} \delta^{4}
\left(p_1 + p_2 - p_3 - p_4 \right)~\frac{1}{4} \sum_{spins} |M|^{2} \nonumber \\
\end{eqnarray}
where $f_{1,2}$, the occupation numbers for the initial colliding electrons, positrons, plasmons or nucleons and $(1 \pm f_{3,4})$, the Pauli blocking factors of the final state bosons ($+$ sign) and fermions($-$ sign).

\section{Radion production in plasmon fusion}

Photons are quite abundant in supernovae. Due to plasma effect inside the supernovae, photons becomes effectively massive. These massive photons are called plasmons. Our interest is in the production of a light stabilized radion $\phi$ in plasmon-plasmon annihilation 
\beq
\gamma_P(p_1) + \gamma_P(p_2) \to \phi(k).
\eeq 
The interaction vertex of the plasmon-plasmon-radion $\gamma_P(p_1)-\gamma_P(p_2)-\phi(k)$ is given by \cite{GW}
\beq
- \frac{2 i m_A^2}{\langle \phi \rangle} \eta^{\mu \nu}
\eeq

\noindent In the c.o.m frame, the 4-momentum vectors associated with the incoming and outgoing particles are given by 
\bea
p_1^{\mu}&=& (E_1, 0, 0, p),~p_2^{\mu} = (E_2, 0, 0, -p),\\
k^{\mu} &=& (E_\phi, 0, 0, 0).
\eea

\noindent where $p = |\vec{p}|$. It  often turns out to be more convenient to keep the polarizations explicitly. Here $\epsilon_{\mu}^{\pm}$ and $\epsilon_{\mu}^{0}$ are the transverse and longitudinal polarization vectors of a massive gauge boson. For a massive vector
boson(\eg ~plasmon) with momentum $k^{\mu}=(E,0,0,k)$ and mass $m_A$, the components of the polarization vector can be written as 
\bea
 \epsilon^+_{\mu}(k)&=&\frac{1}{\sqrt{2}}(0,1,i,0)\ ,\\
 \epsilon^-_{\mu}(k)&=&\frac{1}{\sqrt{2}}(0,-1,i,0)\ ,\\
 \epsilon^0_{\mu}(k)&=&\frac{1}{m_A}(k,0,0,-E)\ .
\eea
The polarization vectors satisfy the following normalization and polarization sum conditions
\bea
e^{s\,\mu}e^{s'\, *}_{\mu} = 4\delta^{s s'},~~
\sum_{s=1}^3 e^s_{\mu}(k)e^{s\, *}_{\nu}(k) = -\eta_{\mu\nu} + \frac{k_\mu k_\nu}{m_A^2}, ~~e^{s\,\mu\nu}e^{s'\, *}_{\mu\nu} = 4\delta^{s s'}\ .
\eea

\noindent The total squared amplitude, averaged over the initial three
polarizations(massive plasmons have three state of polarizations) states and summed over final states for the process 
$\gamma_P(k_1)+\gamma_P(k_2) \to \phi(p)$ is
\bea \label{Eq:amplsq}
\overline{|M|^{2}}  =  \left(\frac{1}{3}\right)^{2} \left(\frac{2 \om_{0}^{2}}{\langle \phi \rangle}\right)^{2} \left\lbrace 2 + \left( 1-\frac{s}{2 \om_{0}^{2}} \right)^{2} \right\rbrace 
\eea

\noindent Sunstituting this in Eq.~\ref{2to1crossgen}, we find the total cross-section $\sigma_{\g_P \g_P \rightarrow \phi}$ as   
\bea 
\sigma_{\gamma_{p}\gamma_{p}\rightarrow \phi} & = & \frac{1}{2s} \overline{|M|^{2}} \left(2\pi\right)^{4}\int \frac{d^{3}\bar{p}}{2(2\pi)^{3}E_{\phi}}\delta^{4}\left(p-(k_{1}+k_{2})\right) \nonumber  \\
& = & \frac{1}{2s} \left(\frac{1}{3}\right)^{2} \left(\frac{2 \om_{0}^{2}}{\langle \phi \rangle}\right)^{2}\left\lbrace 2 + \left( 1-\frac{s}{2 \om_{0}^{2}} \right)^{2} \right\rbrace 2 \pi \delta\left(s-m_{\phi}^{2}\right) 
\label{2to1totcross}
\eea
The volume emissivity of a supernova  with a temperature $T$  through this process is obtained by thermal-averaging over the Bose-Einstein distribution. Hence, the energy loss rate  
($\dot{\varepsilon}_{\g_P} = \frac{1}{\rho_{SN}}\dot{Q}_{\g_P}$) due to plasmon plasmon annihilation is given by (similar to that of the energy loss rate via 
$\gamma\gamma \to \nu\bar{\nu}$. \cite{ggnn})
\bea 
\dot{\varepsilon}_{\g_{p}\g_{p}\rightarrow \phi} & = & \frac{1}{\rho_{SN}} \langle n_{\g_{p}} n_{\gamma_{p}} \sigma_{\g_{p}\g_{p}\rightarrow \phi} V_{rel} E_{c.m}\rangle \nonumber \\
 &=& \frac{1}{\rho_{SN}} \frac{1}{\pi^4}
\int_{\om_0}^{\infty} d \om_1 \frac{\om_1 (\om_1^2 - \om_0^2)^{1/2}}{e^{\om_1/T}-1}  \int_{\om_0}^{\infty}  d \om_2 \frac{\om_2 (\om_2^2 - \om_0^2)^{1/2}}{e^{\om_2/T}-1}
~{S (\om_1+\om_2)\over 2\om_1\om_2}~  \s_{\g_{p}\g_{p}\rightarrow \phi}, 
\eea 
where $\s_{\g_{p}\g_{p}\rightarrow \phi}$ is given in Eq.~\ref{2to1totcross}. 
Note that $N_{\g_P}= \frac{1}{\pi^2}
\int_{\om_0}^{\infty} d \om \frac{\om (\om^2 - \om_0^2)^{1/2}}{e^{\om/T}-1} $ is the number density of thermal photons or plasmons. In the present case, we treat the plasmon to be transverse (with the dispersion relation given by  $\om^2 = \om_0^2 + |\bf{k}|^2$), since the contribution coming from the longitudinal plasmon is typically smaller \cite{Canuto,PR}. In above $\om_0$ corresponds to plasma frequency in the supernovae core.

Finally introducing the dimensionless variables $x_i = \om_i/T$($i=0,1,2$) and performing the $x_2$ integration, we find the reaction rate as  
\bea \label{Eq:lossrate}
\dot{\varepsilon}_{\g_P} = \frac{T^7 x_0^4}{9 \pi^3 \vphi^{2}}  \frac{1}{\rho_{SN}} \mathcal{F}
 \int_{x_0}^{\infty} d x_1 \frac{(x_1^2 - x_0^2)^{1/2}}{e^{x_1}-1} \frac{\left[(\frac{m_\phi}{T} - x_1)^2 - x_0^2 \right]^{1/2}}{e^{\frac{m_\phi}{T} - x_1}-1},
\eea
where $\mathcal{F} = \left[3 - \left(\frac{m_\phi}{x_0 T}\right)^2 + \frac{1}{4} \left(\frac{m_\phi}{x_0 T}\right)^4\right]$. In above we have used $\delta \left[s - m_\phi^2 \right] = \frac{1}{T^2} \delta\left[(x_1 + x_2)^2 - \frac{m_\phi^2}{T^2}\right]$ (while doing the $x_2$ integration) where $s = (\om_1 + \om_2)^2 = T^2 (x_1 + x_2)^2$. Also $\om_0 = T x_0$ where $\om_0$ is taken to be equal to $m_A$ (the transverse plasmon mass). 

\section{Neutrino pair production in electron-positron collision}
The neutrino pair production via the  $s$-channel exchange of a light stabilized radion 
$\phi$ produced in electron($e^-$)-positron($e^+$) collision may play a crucial role in the SN1987A cooling. The process is  
\beq
e^{-}(p_1) + e^{+}(p_2)  \stackrel{\phi}{\longrightarrow}   \nu_l (p_3) + \overline{\nu_l}(p_4).
\eeq 
The radion-fermion-fermion $f(p_1) - \overline{f}(p_2) - \phi(k)$ is given by \cite{GW}
\beq
- \frac{3 i}{2\vphi} \left[\slashed{p_1} - \slashed{p_2} - \frac{8}{3} m_f\right]
\eeq
where $p_1,~p_2$ are the incoming momenta 
In the c.o.m frame, the momentum vectors for this reactions are
\bea
p_1^{\mu}&=& (E_1, 0, 0, |\vec{p}|),~p_2^{\mu}= (E_2, 0, 0, -|\vec{p}|),\\
p_3^{\mu}&=& (E_3, 0, 0, |\vec{p'}|),~ p_4^{\mu}= (E_4, 0, 0, -|\vec{p'}|).
\eea 

The total squared amplitude (averaged over the initial spin states and summed over final spin states) for the process is
\bea \label{Eq:amplsq}
\overline{|M|^{2}}  = \frac{1}{4} \sum_{spins} |M|^{2} =  \left(\frac{1}{2}\right)^{2} \left(\frac{4 m_{e}^{2} m_{\nu_l}^2}{{\langle \phi \rangle}^4}\right) \frac{(s-4 m_{e}^2)(s-4 m_{\nu_l}^2)}{\left[(s-m_\phi^2)^2 + m_\phi^2 \Gamma_\phi^2\right]} 
\eea

Finally, the volume energy-loss rate per unit mass of a supernovae at a temperature $T$ is given by  
\begin{eqnarray}
 \dot{\varepsilon} & = & \frac{1}{\rho_{SN}} \prod^{4}_{i=1} \int \frac{ d^{3}\textbf{p}_{i} }{2E_{i}(2 \pi)^{3}} f_1 f_2 (1 - f_3) (1 - f_4)~(2 \pi)^{4} \delta^{4}
\left(p_1 + p_2 - p_3 - p_4\right)~ \overline{|M|^{2}}
\end{eqnarray}
which after some rearranging can be written as  
\begin{eqnarray}
 \dot{\varepsilon} = \frac{T^{7}}{64 \pi^{5} \rho_{SN}} & * & \iint \frac{x_{1}x_{2}}{\left[ 1 + exp(x_{1}- \frac{\mu_{e^{-}}}{T})  \right] \left[ 1 + exp(x_{2}+ \frac{\mu_{e^{-}}}{T}) \right] }  \nonumber \\
 & * & \frac{exp(x_{1}+x_{2}) dx_{1} dx_{2}}{\left[ 1 + exp(\frac{x_{1}+x_{2}}{2}+ \frac{\mu_{\nu_{l}}}{T}) \right]  \left[ 1 + exp(\frac{x_{1}+x_{2}}{2}-\frac{\mu_{\nu_{l}}}{T})  \right]}  \overline{|M|^{2}} 
 \end{eqnarray}
where 
\begin{eqnarray*}
 \overline{|M|^{2}} & = & \frac{ m_{\nu_{l}}^{2}m_{e}^{2}}{ T^{3}\langle \phi \rangle^{4} \left[ (T^2 (x_1 + x_2)^2 - m_{\phi}^{2})^{2} + m_{\phi}^2 \Gamma_{\phi}^2 \right] } \\
& \times & \left[ T^4 (x_1 + x_2)^4 - 4 T^2 (x_1 + x_2)^2 (m_{e}^{2}+m_{\nu_{l}}^{2}) + 16 m_{e}^{2}m_{\nu_{l}}^{2} \right]
\end{eqnarray*}
Here $s = (E_1 + E_2)^2 = T^2 (x_1 + x_2)^2$ and $\om_0$ is defined above.  
\section{Neutrino pair production in plasmon-plasmon collision}
The third process that we are interested to look at is the neutrino pair production in plasmon-plasmon collision via the $s$-channel exchange of a light stabilized radion $\phi$ \ie  
\beq
\gamma_P(k_1) + \gamma_P(k_2) \stackrel{\phi}{\longrightarrow}   \nu_l (p_1) + \overline{\nu_l}(p_2).
\eeq 
The total squared amplitude (averaged over the initial polarization states and summed over the final spin states) is given by 
\bea \label{Eq:amplsq}
\overline{|M|^{2}}  = \left(\frac{1}{3}\right)^2 \sum_{spins} |M|^{2} =  \frac{16 m_{A}^{2} m_{\nu_l}^2}{{9 \langle \phi \rangle}^4} \frac{(s-4 m_{\nu_l}^2)}{\left[(s-m_\phi^2)^2 + m_\phi^2 \Gamma_\phi^2\right]} \left[1 + \frac{1}{2}\left(1 - \frac{s}{2 m_A^2}\right)^2\right] 
\eea

\noindent With this the energy loss rate (volume emissivity)  $Q$ due to this process can be written as  
\begin{eqnarray}
 Q & = & \prod^{2}_{i=1} \int \frac{ 2^2  d^{3}\textbf{k}_{i} }{2E_{i}(2 \pi)^{3}} 
\prod^{4}_{i=3} \int \frac{ d^{3}\textbf{p}_{i} }{2E_{i}(2 \pi)^{3}} f_1 f_2 (1 - f_3) (1 - f_4)~ \nonumber \\
& & \times  (2 \pi)^{4} \delta^{4}
\left(p_1 + p_2 - p_3 - p_4\right) \frac{1}{4} \sum_{spins} |M|^{2} 
\end{eqnarray}
\noindent where $f_{1,2}$, the occupation numbers for the initial state plasmons and 
$(1 - f_{3,4})$, the Pauli blocking factors of the final state neutrino and anti-neutrinos. \\
 Finally, the energy loss rate(volume emissivity) per unit mass at temp. $T$ is given by 
\begin{eqnarray}
 \dot{\varepsilon} = \frac{T^{7}}{16 \pi^{5} \rho_{SN}} & * & \iint dx_1 dx_2 \frac{x_{1}x_{2}}{\left[ exp(x_{1}) - 1 \right] \left[ 1 + exp(x_{2}+ \frac{\mu_{e^{-}}}{T}) \right] }  \nonumber \\
 & * & \frac{exp(x_{1}+x_{2}) dx_{1} dx_{2}}{\left[ 1 + exp(\frac{x_{1}+x_{2}}{2}+ \frac{\mu_{\nu_{l}}}{T}) \right]  \left[ 1 + exp(\frac{x_{1}+x_{2}}{2}-\frac{\mu_{\nu_{l}}}{T})  \right]} \overline{|M|^{2}} 
 \end{eqnarray}
where 
\begin{eqnarray*}
 \overline{|M|^{2}}  & = & \frac{16 m_{\nu_{l}}^{2} m_{A}^{4}}{9 T^{3}\langle \phi \rangle^{4} } \frac{\left[T^2 (x_1 + x_2)^2 - 4 m_{\nu_{l}}^{2}\right]}{\left[ (T^2 (x_1 + x_2)^2 - m_{\phi}^{2})^{2} + m_{\phi}^2 \Gamma_{\phi}^2 \right] } \\
& \times & \left[ 1 + \frac{1}{2} \left(1 - \frac{T^2 (x_1 + x_2)^2}{2 m_A^2}\right)^2\right]
\end{eqnarray*}
where $s = (E_1 + E_2)^2 = T^2 (x_1 + x_2)^2$. 

\section{Numerical Analysis}
As discussed above the processes of our interest that may contribute in the SN1987A cooling are radion mediated three processes:  (i)  $\gamma_P + \gamma_P \to \phi$,~
(ii) $e^{-} + e^{+}  \stackrel{\phi}{\longrightarrow}   \nu_x + \overline{\nu_x}$ and (iii) $\gamma_P + \gamma_P \stackrel{\phi}{\longrightarrow}   \nu_x + \overline{\nu_x}$ (with $x=\mu,~\tau$).  If the SN1987A cooling is due to a class of radion mediated processes the emissivity rate for those channels must be $\dot{\varepsilon} \le 10^{19}~{erg~g^{-1}~s^{-1}}$ \cite{Raffelt}, which can be converted to a lower bound on radion vev $\vphi$ as discussed below.    

\subsection{Bound on $\vphi$ from the $\gamma_P + \gamma_P \to \phi$ decay}
\begin{figure}[htbp]
\vspace{5pt}
\centerline{\hspace{-3.3mm}
{\epsfxsize=6cm\epsfbox{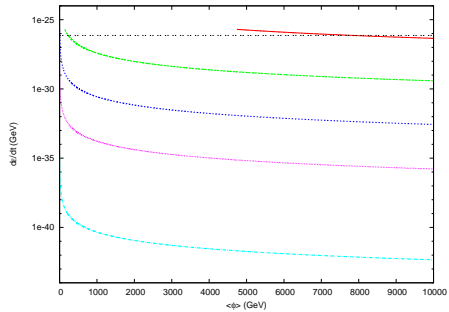}} }
\vspace*{-0.25in}
\caption{\it{The supernovae energy-loss rate $d\varepsilon/dt~\rm{GeV^{-1}}$ due to radion emission produced in plasmon-plasmon annihilation is shown as a function of $\vphi$ (GeV). While going from the topmost curve to the lowermost curve, $m_\phi$ increases as $1.25,~1.5,~1.75,~2.0$ and $2.5$ GeV. The upper horizontal line corresponds to the upper bound of the supernovae energy loss rate i.e. $\dot{\varepsilon} \le 7.288\times 10^{-27} \rm{GeV}$.}}
\protect\label{plpl_rad}
\end{figure}

In Fig. \ref{plpl_rad} we have shown the energy-loss rate $\dot{\varepsilon}(\gamma_P + \gamma_P \to \phi)$ as a function of the radion $\vphi$ corresponding to an ultra-light stabilized radion of mass $m_\phi = 1.25,~1.5,~1.75,~2.0$ and $2.5$ GeV. 
The upper horizontal line corresponds to the upper bound of the supernovae energy loss rate $\dot{\varepsilon} = 7.288\times 10^{-27} \rm{GeV}$ for any new physics channel. From the topmost curve to the lowermost curve the radion mass $m_\phi$ increases as $1.25,~1.5,~1.75,~2.0$ and $2.5$ GeV. We find $\vphi > 7852~\rm{GeV}$ for $m_\phi = 1.25~\rm{GeV}$ and  $\vphi > 230~\rm{GeV}$ for $m_\phi = 1.5~\rm{GeV}$. No bound follows for $m_\phi = 1.75,~2.0$ and $2.5$ GeV.
\subsection{Bound on $\vphi$ from the $e^{-} + e^{+}  \stackrel{\phi}{\longrightarrow}   \nu_l + \overline{\nu_l}$  scattering }
\vspace{-0.15in}
\noindent The muon and tau neutrino or anti-neutrino produced in $e^- ~e^+$ annihilation (due to a  $s$ channel exchange of a light stabilized radion) may take away the supernovae energy and hence can explain the deficit of electron type of (anti)neutrino at the detector recorded by the Kamiokande and IMB collaborations. 
\begin{figure}[htbp]
\centerline{\hspace{-3.3mm}
{\epsfxsize=6.0cm\epsfbox{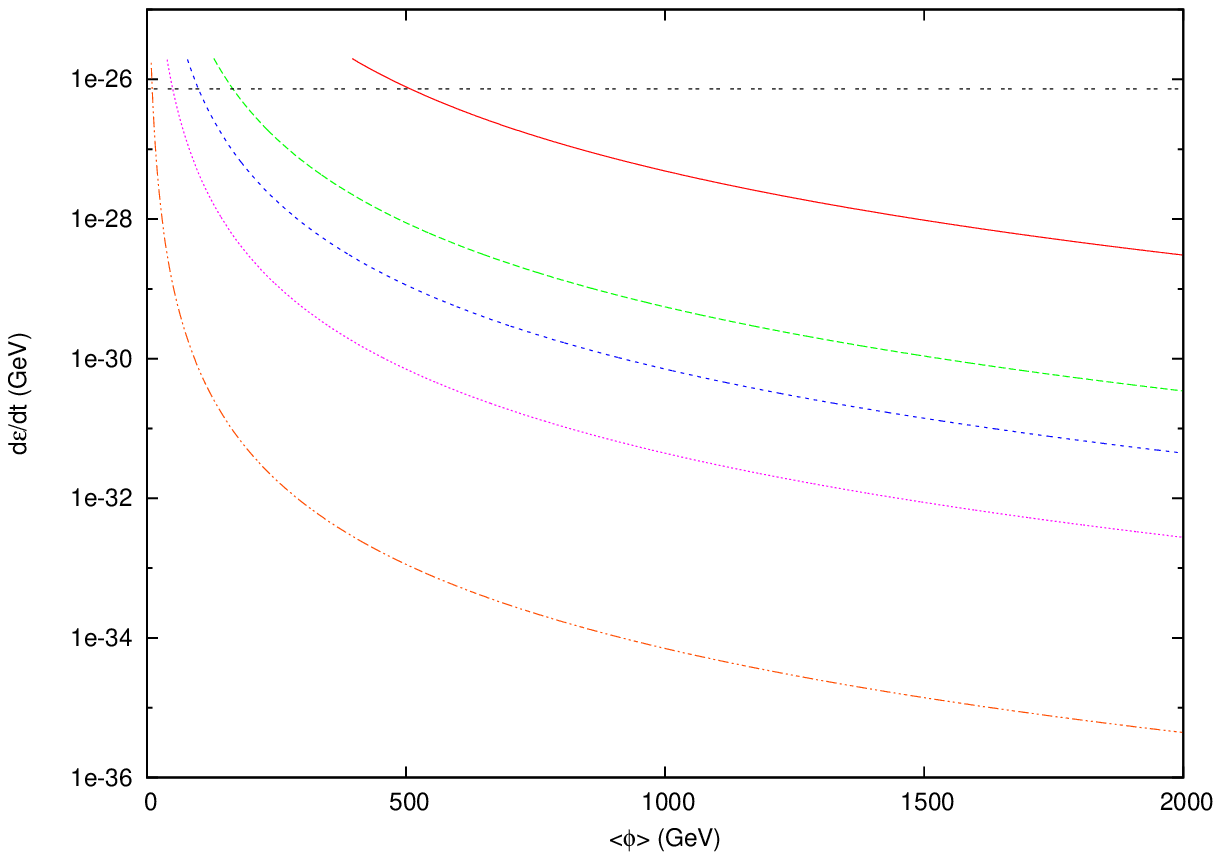}} \hspace{0.15in} {\epsfxsize=6.0cm\epsfbox{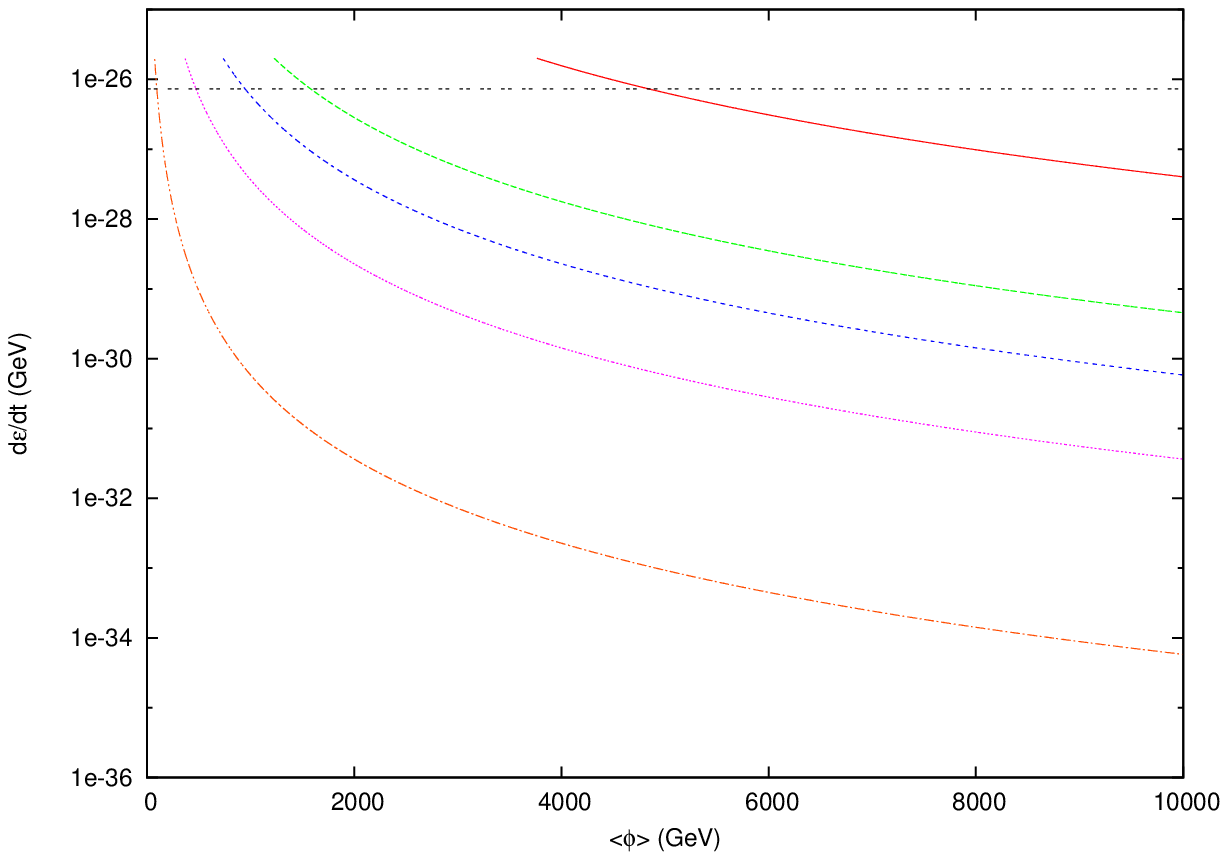}} }
\vspace*{-0.25in}
\caption{\it{The energy-loss rate $d\varepsilon/dt~\rm{GeV^{-1}}$ is shown as a function of $\vphi$ (GeV). From the topmost to the lowermost curve, $m_\phi$ increases as $1,~3,~5,~10$ and $50$ GeV, respectively. The horizontal line corresponds to $\dot{\varepsilon} \le 7.288\times 10^{-27} \rm{GeV}$. The left panel corresponds to the energy loss due to $e^{-} + e^{+}  \stackrel{\phi}{\longrightarrow}   \nu_\mu + \overline{\nu_\mu}$  scattering, whereas the right panel corresponds to the energy loss by $e^{-} + e^{+}  \stackrel{\phi}{\longrightarrow}   \nu_\tau + \overline{\nu_\tau}$  scattering. }}
\protect\label{emep_rad_nunubar}
\end{figure}
On the left panel of Fig.~\ref{emep_rad_nunubar} we have shown the energy loss rate $\dot{\varepsilon}(e^{-} + e^{+}  \stackrel{\phi}{\longrightarrow}   \nu_\mu + \overline{\nu_\mu})$ as a function of $\vphi$, whereas on the right panel we have shown the energy loss rate  $\dot{\varepsilon} (e^{-} + e^{+}  \stackrel{\phi}{\longrightarrow}   \nu_\tau + \overline{\nu_\tau})$  as a function of $\vphi$.  
 The horizontal line on each panel corresponds to the upper bound $\dot{\varepsilon} \le 7.288\times 10^{-27} \rm{GeV}$. On the left(right) panel, from the topmost to the lowermost curves $m_\phi$ increases as $1,~3,~5,~10$ and $50$ GeV. 
\begin{figure}[htbp]
\vspace{-3.0pt}
\centerline{\hspace{-3.3mm}
{\epsfxsize=6.0cm\epsfbox{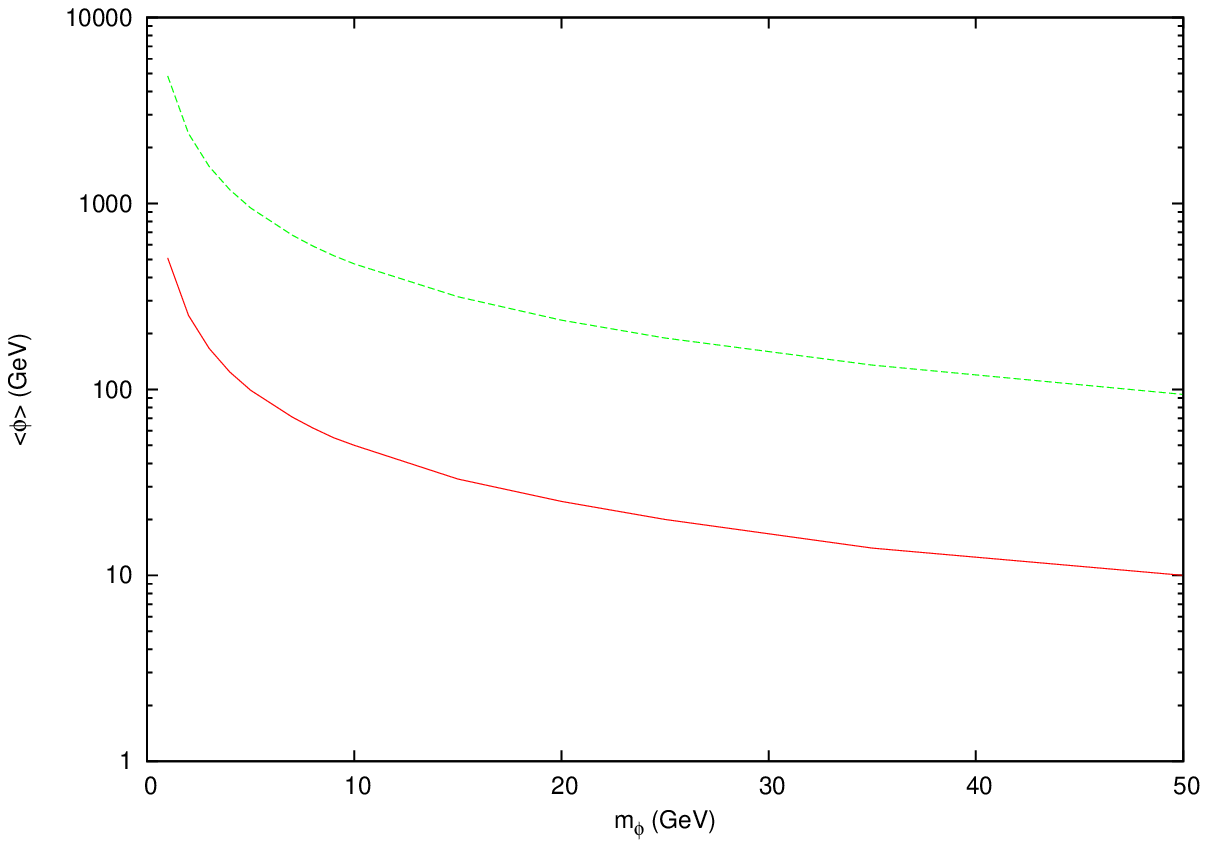}} }
\vspace*{-0.25in}
\caption{\it{The lower bound on $\vphi$ is plotted against $m_\phi$. The lower and the upper curves follows from the fact that the supernovae cooling stems from the $e^{-} + e^{+}  \stackrel{\phi}{\longrightarrow}  \nu_\mu  + \overline{\nu_\mu}$ and $e^{-} + e^{+}  \stackrel{\phi}{\longrightarrow}  \nu_\tau + \overline{\nu_\tau}$ scattering processes, respectively. The region above each curve is allowed. }}
\protect\label{mRvR_emepnu}
\end{figure}

In Fig. \ref{mRvR_emepnu}, we have shown $m_\phi$ as a function of $\vphi$ 
(the lower bound). The bound follows from the fact if any new physics channel contribute to the SN1987A energy loss, it should be $\dot{\varepsilon} \le 7.288\times 10^{-27} \rm{GeV}$. The lower(upper) curve follows from the fact that the SN1987A cooling is caused by the 
$e^{-} + e^{+}  \stackrel{\phi}{\longrightarrow}  \nu_\mu (\nu_\tau) + \overline{\nu_\mu}(\overline{\nu_\tau})$  radion mediated processes. The region above each curve is allowed.

 Below in Table 2 we have shown the lower bound on $\vphi$ for different $m_\phi$ (refer Fig. \ref{mRvR_emepnu}).  
From Table 2 we see that for a given channel, the lower bound on $\vphi$ decreasees with the increase in $m_\phi$. Also for a particular $m_\phi$ the lower bound on $\vphi$ (third column) which follows from $\nu_\tau - \overline{\nu_\tau}$ production is much higher than that(second column) follows from  $\nu_\mu - \overline{\nu_\mu}$ production. Note that the bound on $\vphi$ (second column) is much weaker  for a radion of mass of few tens of a GeV.
\vspace*{-0.2in}
\begin{center}
Table 2
\end{center}
\begin{center}
\begin{tabular}{|c|c|c|c|}
\hline
$m_\phi$ (GeV) & $\vphi$ (GeV) (lower curve) & $\vphi$ (GeV) (upper curve)  \\
\hline
\hline
  1  &  509 &  4846\\
  3  &  166 & 1580\\
  5   &  99 & 946\\
  10  &  50 &  473\\
  15  &  33 &  315\\
  20  &  25 &  236\\
  25  &  20 &  189\\
\hline 
\end{tabular}
\end{center}
\noindent {\it Table 2: The lower bound on $\vphi$(follows from Fig. \ref{mRvR_emepnu}) corresponding to different $m_\phi$ is shown.  The second(third) column follows from the energy loss rate $\dot{\varepsilon}(e^{-} + e^{+}  \stackrel{\phi}{\longrightarrow}  \nu_{\mu(\tau)} + \overline{\nu}_{\mu(\tau)}) \le 7.288\times 10^{-27} \rm{GeV}$.}

\subsection{Bound on $\vphi$ from the $\gamma_P + \gamma_P  \stackrel{\phi}{\longrightarrow}   \nu_l + \overline{\nu_l}$  scattering }
\begin{figure}[htbp]
\vspace{-0.1in}
\centerline{\hspace{-3.3mm}
{\epsfxsize=6cm\epsfbox{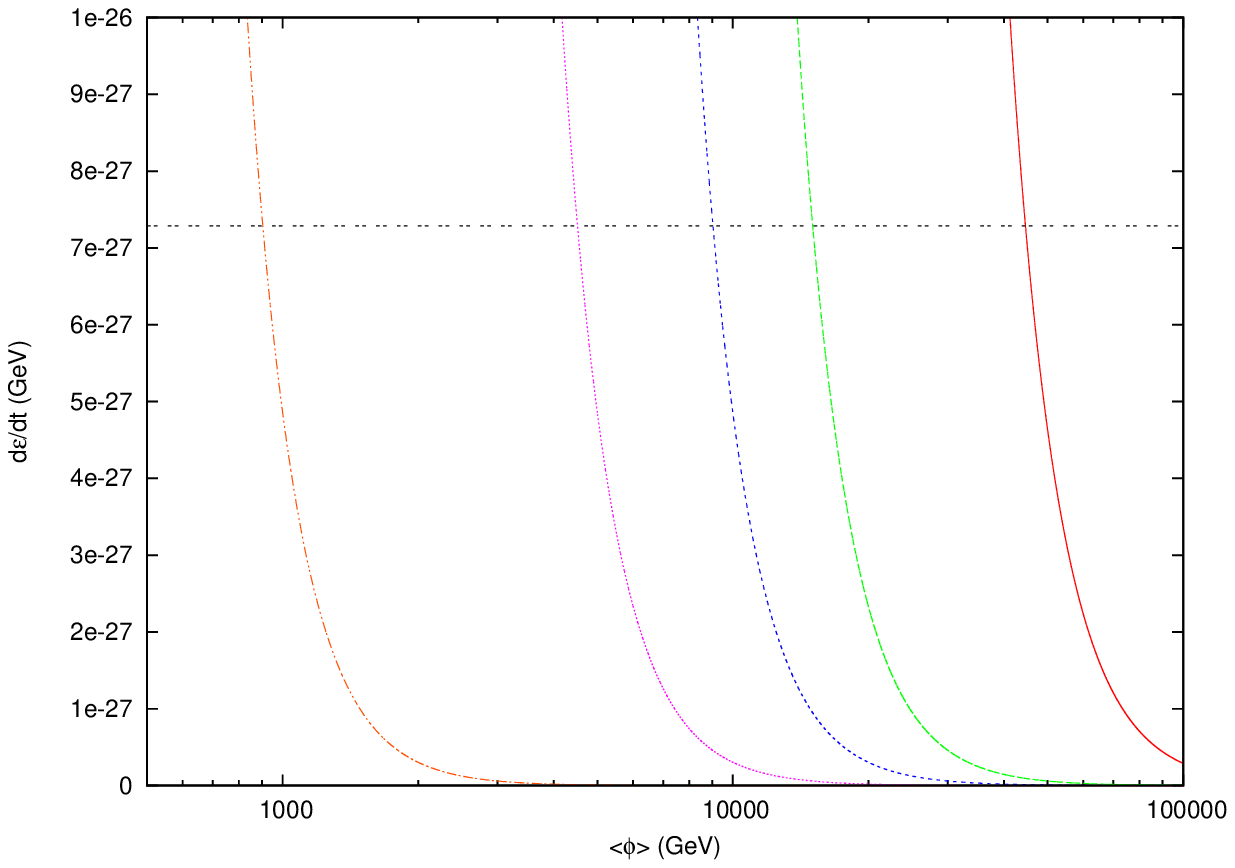}} \hspace{0.15in} {\epsfxsize=6cm\epsfbox{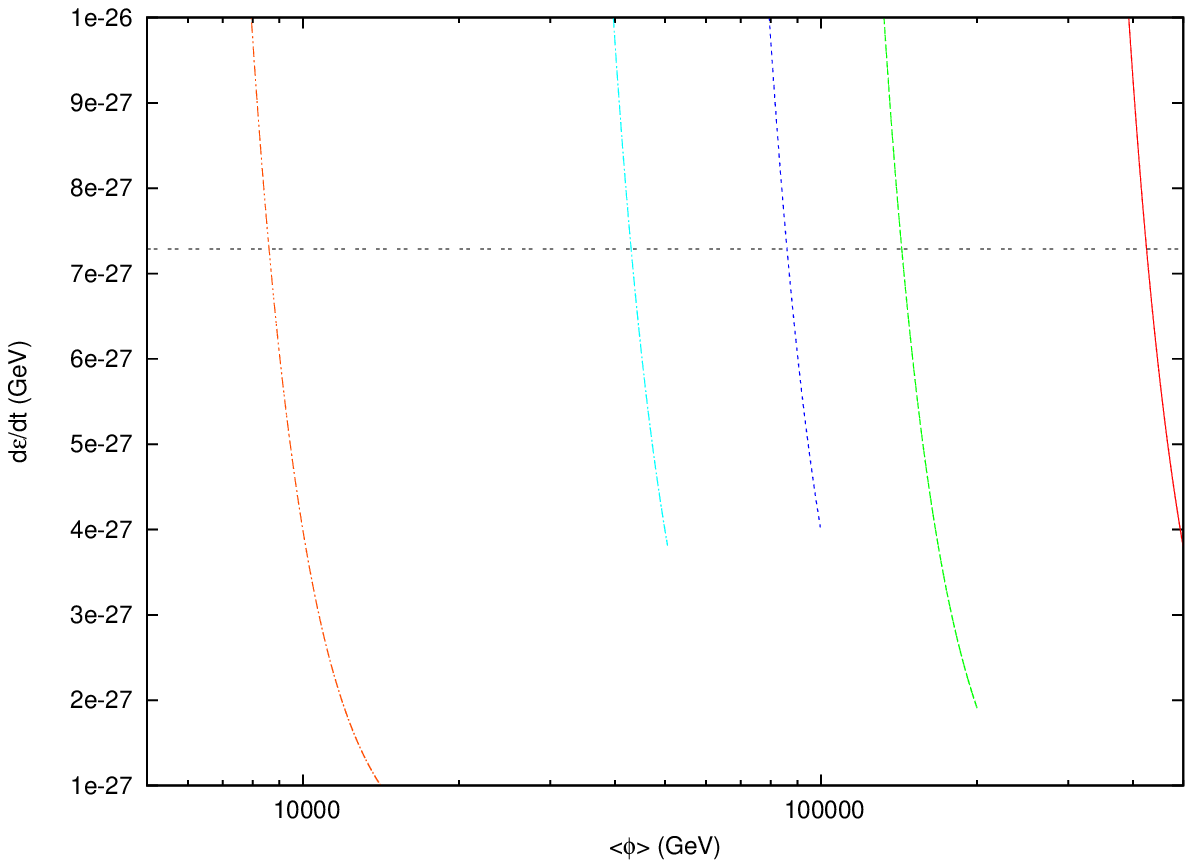}} }
\vspace*{-0.25in}
\caption{\it{The energy-loss rate $d\varepsilon/dt~\rm{GeV^{-1}}$ is shown as a function of $\vphi$ (GeV). From the rightmost to the leftmost curve, $m_\phi$ increases as $1,~3,~5,~10$ and $50$ GeV, respectively. The horizontal line corresponds to the upper bound on the energy loss rate i.e. $\dot{\varepsilon} \le 7.288\times 10^{-27} \rm{GeV}$. The left(right) panel corresponds to the energy loss by $\gamma_P + \gamma_P  \stackrel{\phi}{\longrightarrow}   \nu_\mu + \overline{\nu_\mu}$ ($\gamma_P + \gamma_P  \stackrel{\phi}{\longrightarrow}   \nu_\tau + \overline{\nu_\tau}$) processes.}}
\protect\label{plpl_rad_nunubar}
\end{figure}

\begin{figure}[htbp]
\centerline{\hspace{-3.3mm}
{\epsfxsize=6cm\epsfbox{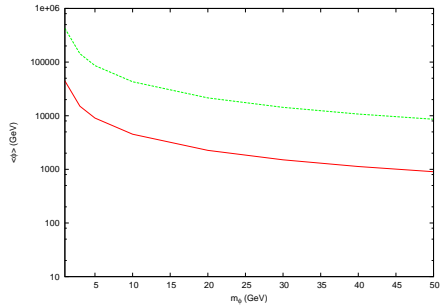}} }
\vspace*{-0.15in}
\caption{\it{ The lower bound on $\vphi$ is plotted against $m_\phi$.  The lower bound follows from the fact for any new physics channel contributing to the SN1987A energy loss, the loss rate $\dot{\varepsilon} \le 7.288\times 10^{-27} \rm{GeV}$. The lower(upper) curve follows from the fact that the processes $\gamma_P + \gamma_P \stackrel{\phi}{\longrightarrow}  \nu_\mu (\nu_\tau) + \overline{\nu_\mu}(\overline{\nu_\tau})$ contribute to the SN1987A cooling. The region above each curve is allowed. }}
\protect\label{mRvR_plplnu}
\end{figure}

\noindent The Supernovae may cools off by emitting muon and tau neutrino and anti- neutrino produced in plasmon-plasmon annihilation inside it's core. The relevant processes are  $\gamma_P + \gamma_P  \stackrel{\phi}{\longrightarrow}   \nu_x + \overline{\nu_x}$ (with $x = \mu,~\tau$). In Fig. \ref{plpl_rad_nunubar}, we have shown the energy-loss rate as a function of $\vphi$ corresponding to different $m_\phi$ values.
With the lowering of radion mass, the lower bound on $\vphi$ decreases. In Fig. \ref{mRvR_plplnu}, we have plotted $\vphi$ (the lower bound) as a function of radion mass $m_\phi$. 
The lower and the upper curves correspond to the SN1987A cooling due to $\gamma_P + \gamma_P  \stackrel{\phi}{\longrightarrow}  \nu_\mu + \overline{\nu_\mu}$ and $\gamma_P + \gamma_P  \stackrel{\phi}{\longrightarrow}   \nu_\tau + \overline{\nu_\tau}$ processes, respectively. 
In Table 3 we have shown the lower bound on $\vphi$ for different $m_\phi$ values.  For an ultra-light radion of mass $1$ GeV, we find the lower bound on $\vphi$ as $44.6 ~\rm{TeV}$ and $\sim 425~\rm{TeV}$, respectively whereas for a radion of mass about $50$ GeV, we find the lower bound as $0.9$ TeV and $8.6$ TeV, respectively.
\begin{center}
Table 3
\end{center}
\vspace*{-0.2in}
\begin{center}
\begin{tabular}{|c|c|c|c|}
\hline
$m_\phi$ (GeV) & $\vphi$ (GeV) (left panel) & $\vphi$ (GeV) (right panel)  \\
\hline
\hline
  1   & 44648 & 424951\\
  3   & 15042 & 143147 \\
  5   & 9033  & 85958 \\
  10  & 4518  & 42993  \\
  20  & 2259  & 21498  \\
  30  & 1506  & 14332\\
  50  & 904   & 8600  \\
\hline 
\end{tabular}
\end{center}
\noindent {\it Table 3: The lower bound on $\vphi$ as a function of $m_\phi$ corresponding to the SN1987A energy loss rate $\dot{\varepsilon} \le 7.288\times 10^{-27} \rm{GeV}$. The second(third) column of the lower bound on $\vphi$ follows from the energy loss via the channel $\gamma_P + \gamma_P  \stackrel{\phi}{\longrightarrow}  \nu_{\mu(\tau)} + \overline{\nu}_{\mu(\tau)}$.}

\section{Conclusion}
In this work we have investigated the impact of a light stabilized radion in supernovae cooling. The radion produced inside the core of a supernova due to electron-positron and plasmon-plasmon collisions, can take away much of the energy released in supernovae explosion. The primary processes of our concern are (i) $\gamma_P + \gamma_P  \to \phi$,~(ii) $e^+ + e^- \stackrel{\phi}{\longrightarrow}  \nu_{\mu(\tau)} + \overline{\nu}_{\mu(\tau)}$and (iii) $\gamma_P + \gamma_P  \stackrel{\phi}{\longrightarrow}  \nu_{\mu(\tau)} + \overline{\nu}_{\mu(\tau)}$. Assuming that the energy loss rate due to each of the above three channels $\dot{\varepsilon} \leq 7.288\times 10^{-27} \rm{GeV}$, we obtain the following lower bound on the radion vev $\vphi$ due to a light stabilized radion. For the process $\gamma_P + \gamma_P  \to \phi$: we find $\vphi \geq 7.85~\rm{TeV}$ for $m_\phi = 1.25~\rm{GeV}$ and  $\vphi \geq 2.3~\rm{GeV}$ for $m_\phi = 1.5~\rm{GeV}$. For $e^+ + e^- \stackrel{\phi}{\longrightarrow}  \nu_{x} + \overline{\nu}_{x}$:  with $x= \tau$ we find $\vphi  \geq  0.95$ TeV and $\geq 0.24$ TeV corresponding to $m_\phi = 5 $ GeV and $20$ GeV, respectively. No reasonable bound on $\vphi$ follows with $\nu_{\mu},\overline{\nu}_{\mu}$ as final state particles. Finally the process $\gamma_P + \gamma_P  \stackrel{\phi}{\longrightarrow}  \nu_{x} + \overline{\nu}_{x}$: for $x=\mu$, we find $\vphi  \geq 9.0$ TeV and $\geq 2.4$ TeV for $m_\phi = 5 $ GeV and $20$ GeV, respectively. For $x=\tau$, we find $\vphi  \geq  86$ TeV and $\geq 21.5$ TeV corresponding to the same $m_\phi$ values. 
\begin{acknowledgments}
This work is supported by the DAE BRNS Project Ref.No. 2011/37P/08/BRNS and the BITS SEED Grant Project.  
\end{acknowledgments}


\end{document}